\documentclass[final,5p,times,twocolumn,authoryear]{elsarticle}
\usepackage{amssymb}
\usepackage{amsmath}
\usepackage{amsmath}
\usepackage{url}
\usepackage{array}
\usepackage{pifont}
\usepackage[dvipsnames,svgnames]{xcolor}
\usepackage{ifthen}

 \usepackage[
]{sidecap}   
\sidecaptionvpos{figure}{t} 
\renewenvironment{SCfigure}[1][t]{\begin{figure}[#1]}{\end{figure}}

\newcommand{\oneapi}{oneAPI}
\newcommand{\dpecho}{\texttt{DPEcho}}
\newcommand{\acpp}{\texttt{acpp}}
\newcommand{\hpl}{\texttt{HPL}}
\newcommand{\ack}{\section*{Acknowledgments} }
\newcommand{\discl}{ }

\journal{Astronomy $\&$ Computing}

\setboolean{doubleblind}{false}   

\begin{document}
\begin{frontmatter}


\title{SYCL for 
Energy-Efficient Numerical Astrophysics: the case of DPEcho}

\author[lrz]{Salvatore Cielo}
\author[intel]{Alexander Pöppl}
\author[lrz]{Ivan Pribec}

\affiliation[lrz]{organization={Leibniz Supercomputing Centre of the Bavarian Academy of Science and Humanities},
            addressline={Boltzmannstr. 1}, 
            city={Garching bei München},
            postcode={85748}, 
            state={Bavaria},
            country={Germany}}
\affiliation[intel]{organization={Intel Corporation},
            addressline={2200 Mission College Blvd.}, 
            city={Santa Clara},
            postcode={95052}, 
            state={CA},
            country={USA}}


\begin{abstract}
Energy awareness and efficiency policies are gaining more attention, over pure performance (time-to-solution) Key Performance Indicators (KPIs) when comparing the possibilities offered by accelerated systems. But in a field such as numerical astrophysics, which is struggling with code refactorings for GPUs, viable porting paths have to be shown before first. After summarizing the status and recurring problems of astrophysical code accelerations, we highlight how the field would benefit from portable, vendor-agnostic GPU portings.
We then employ the DPEcho SYCL benchmark to compare raw performance and energy efficiency for heterogeneous hardware on a realistic application, with the goal of helping computational astrophysicists and HPC providers make informed decisions on the most suitable hardware. 
Aside from GPUs showing higher efficiency, we argue on the more informative nature of energy-aware KPIs, in that they convey the specific device performance in a data-driven way.
We also present a convenient, flexible and cross-platform energy-measuring pipeline. Finally, we contextualize our results through measures with different compilers, presenting device ("at the cores") versus node ("at the plug") energy and comparing DPEcho with the High-Performance Linpack (HPL) benchmark.

\end{abstract}


\end{frontmatter}

\section{Introduction: GPUs and energy efficiency for computational Astrophysics} \label{introduction}

Numerical sciences, with Astrophysics on the front line, are facing several important challenges with modern computer architectures: top supercomputers offer increasingly accelerated and performant hardware, which scientists struggle to use, for delays in the (complex and time-consuming) efforts in code refactoring, which should not even be a scientist's main focus.

The current system heterogeneity, while presenting a flourishing panorama for HPC, is adding additional burdens and portability issues: different GPUs require different languages and extensions to make use of their full potential, and in practice programmers are often compelled to choose what hardware to program for, or to maintain different back-ends. 

 \begin{figure*}[ht]
    \centering
    \includegraphics[width=0.9\linewidth]{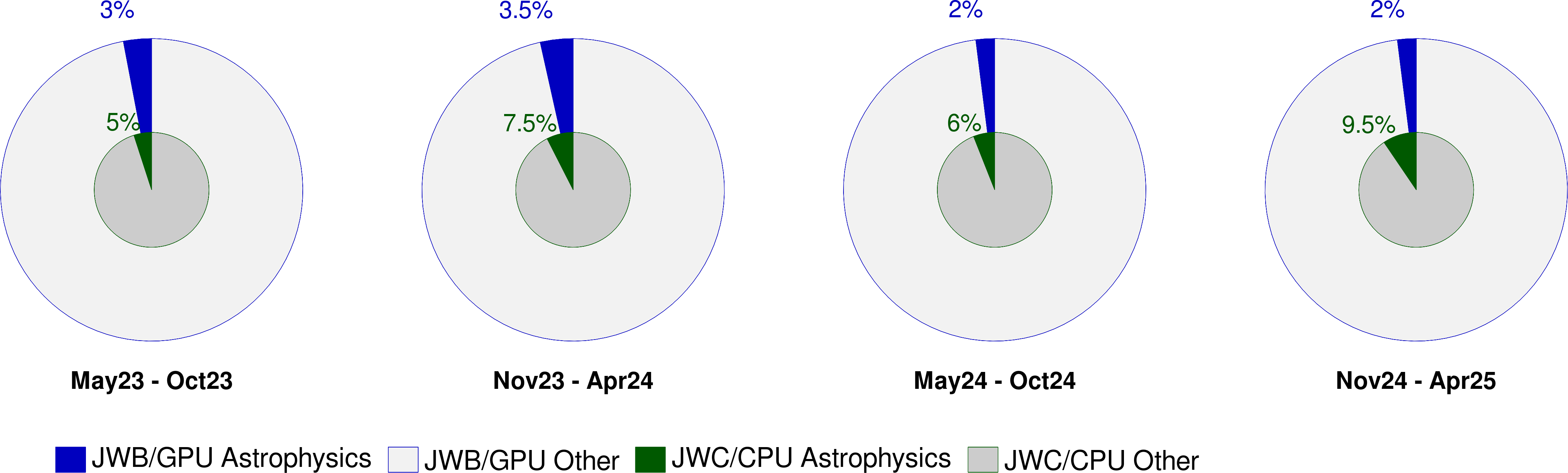}
    \caption{Core-hours percentage allocated for astrophysics projects on the \textit{JUWELS} Supercomputer, at \textit{Jülich Supercomputing Centre}, in the periods from October 2023 to April 2025, for the two partitions \textit{Cluster} (JWC, as a proxy for CPU core-h) and \textit{Booster} (JWB, proxy for GPU core-h).
    The area of the pie charts is proportional to the maximal FLOPs of each partition.
    Data provided by \emph{Gauss Center for Supercomputing}. }    \label{f-alloc}
\end{figure*}

These complications likely result in the observed reduced allocations on top supercomputers\footnote{among other possible factors; e.g. Artificial Intelligence workloads may occupy a larger fraction of GPU allocations (data not available to us). }: Figure \ref{f-alloc} shows the allocation fractions for astrophysics on the supercomputer \emph{Juwels} at the \emph{Jülich Forschungzentrum} (JSC). The charts refer to the \emph{Cluster} and \emph{Booster} partitions, assumed as proxy for CPU and GPU allocations, respectively.  The JSC data for GPUs are the most representative of the whole \emph{Gauss Center for Supercomputing}\footnote{\url{https://www.gauss-centre.eu/}} - one of the largest supercomputing institutions in Europe - as JSC is the centre with the longest GPU tradition and largest user base inside GCS.
In addition, similar results are being presented for other centres while introducing GPU portings of astro- codes \cite{cines2024lesur,idefix2023lesur} and advocating for more efforts in this direction. Clearly computational astrophysics is missing out on the higher FLOPs availability granted by GPU machines, by a factor of about $3$ to $4.75$. While GPU codes are in fact emerging \cite{liska2022harm,peer2023cuharm,begue2023cuharm,kashino2022multi}, likely due to this pressure, the downwards trend worsens with time. Orienting porting efforts towards portable, vendor-agnostic languages such as SYCL would seem rather beneficial in this sense, allowing users a larger device selection.

Meanwhile, serious energetic 
concerns necessitate the adoption of energy-saving or at least energy-aware practices among both vendors and numerical scientists, to the point that energy efficiency is the rationale behind the \emph{Green500} list of supercomputers. 

While several refined Key Performance Indicators (KPIs) related to performance and portability have been suggested in literature (\citealp{pennycook2021portability}),
the adoption of quantities related to energy efficiency is gaining increasing popularity for HPC \cite{dongarra2024hardwaretrendsimpactingfloatingpoint,energyhpc2025germany}, also specifically for computational physics \cite{co-evolutionHPCphys2024} and astrophysics \citep{energy2019taffoni}.  
As we show below, the jump in energy efficiency from CPUs to GPUs is significant. Still, different GPUs may show different efficiencies for a given workload.

 Even for CPUs alone, performance and energy efficiency do not necessarily align \cite{bhac2022cielo}, as high parallelism sometimes comes at the cost of lower CPU frequencies. This matters even more when considering GPUs: the raw performance (in terms of operations per unit time) alone may be misleading, as the size of the devices varies, introducing a measure bias. Astrophysical codes generally exhibit good parallel scaling via MPI,
thus,
what really matters at scale 
is how intrinsically efficient the devices are, rather than how many cores each GPU possesses. Energy efficiency instead does not present such bias, being the ratio of two extensive quantities (i.e. proportional to the device size): number of operations per unit of consumed energy. In other words, consumed energy is an effective, data-driven proxy for device size. 








\section{Methodology}\label{s-perf}

\subsection{DPEcho Setup}\label{ss-setup}

In this work, we present a quantification of the energy efficiency of modern heterogeneous hardware using the \dpecho{} benchmark. 
\dpecho{} is a MPI+SYCL version of the finite-volume General Relativistic Magneto-Hydrodynamics (GR-MHD) code ECHO \cite{echo2007delzanna}, used to model instabilities, turbulence, propagation of waves, stellar winds and magnetospheres, and astrophysical processes around black holes. The code supports classical and relativistic MHD, both in Minkowski or any GR metric (if coded by the user). 
Iterations are done using a 3rd-order Runge-Kutta explicit time stepping on a fixed grid. Along each dimension it uses a 6-point stencil for interpolation and 5-point stencil for reconstruction. For distributed computations, the computational domain is split using a Cartesian MPI decomposition.
In \citealp{dpecho2023cielo} we discussed the code in greater detail.
DPEcho is available publicly\footnote{\ifthenelse{\boolean{doubleblind}}{REDACTED, URL WILL BE FILLED IN.}{\url{https://github.com/LRZ-BADW/DPEcho}}} under a permissive license.

For our benchmark, we choose a standard test problem (same as in \citealp{dpecho2023cielo}): Alfv\`en waves with unit wavenumbers along each direction of a cubic box, thus parallel to the box diagonal, with periodic boundary conditions, full GR-MHD solver, no constrained transport for magnetic field divergence cleaning, highest order reconstruction scheme. 
 \begin{table*}[h!] 
 \renewcommand{\arraystretch}{1.1} 
    \caption[Evaluated Hardware]{Evaluated Hardware, including vendor, model, device type, server and Thermal Design Power (in watt, per single device).}
    \begin{tabular}{ m{0.9cm} c c m{4.0cm} c m{1.2cm} c } 
      \hline
      \centering  
      Alias  & Type & Vendor & Name & Cluster \& Site & TDP & Notes \\
      \hline
      SKX & CPU & Intel & Xeon Platinum 8174 & \ifthenelse{\boolean{doubleblind}}{REDACTED}{SNG-1, LRZ} & 240W & 2 CPUs\\
      SPR & CPU & Intel & Xeon Platinum 8480L & \ifthenelse{\boolean{doubleblind}}{REDACTED}{SNG-2, LRZ} & 350W & 2 CPUs\\
      PVC & GPU & Intel & Data Center GPU Max 1550 & \ifthenelse{\boolean{doubleblind}}{REDACTED}{SNG-2, LRZ} & 450W{*} & 2 Tiles{**} \\
      GNR & CPU & Intel & Xeon 6972P & \ifthenelse{\boolean{doubleblind}}{REDACTED}{Intel} & 500W & 2 CPUs\\
      SRF & CPU & Intel & Xeon 6780E & \ifthenelse{\boolean{doubleblind}}{REDACTED}{Intel} & 330W & 2 CPUs\\
      EMR & CPU & Intel & Xeon Platinum 8592+ & \ifthenelse{\boolean{doubleblind}}{REDACTED}{Intel} & 350W & 2 CPUs \\      
      Rome & CPU & AMD & EPYC 7742 64-Core & \ifthenelse{\boolean{doubleblind}}{REDACTED}{BEAST, LRZ} & 225W & 2 CPUs\\
      Milan & CPU & AMD & EPYC 7773X 64-Core & \ifthenelse{\boolean{doubleblind}}{REDACTED}{BEAST, LRZ} & 280W & 2 CPUs\\
      MI100 & GPU & AMD & Instinct MI100 & \ifthenelse{\boolean{doubleblind}}{REDACTED}{BEAST, LRZ} & 300W & PCIe4  \\
      MI210 & GPU & AMD & Instinct MI210 & \ifthenelse{\boolean{doubleblind}}{REDACTED}{BEAST, LRZ} & 300W & PCIe4\\
      A40 & GPU & NVIDIA & A40 40GB DDR6 & \ifthenelse{\boolean{doubleblind}}{REDACTED}{Alex, NHR@FAU} & 300W \\
      A100 & GPU & NVIDIA & A100 40GB HBM2 & \ifthenelse{\boolean{doubleblind}}{REDACTED}{Alex, NHR@FAU} & 400W\\
      H100 & GPU & NVIDIA & H100 & \ifthenelse{\boolean{doubleblind}}{REDACTED}{Intel} & 350W & PCIe\\
      \hline
    \end{tabular}  
    {  
         \footnotesize 
         \\ {*}\footnotesize{Power-capped to 450W at \ifthenelse{\boolean{doubleblind}}{REDACTED}{LRZ}, nominally 600W}
         \\ {**}\footnotesize{Presented results are normalized to a single tile}
    }  
    \label{t-dev} 
\end{table*}
The problem size and the number of MPI ranks vary depending on the device (see Table \ref{t-dev} for the full list of tested hardware), in order to use the largest part of the available node memory, and use all devices in a single node with one rank per device or NUMA domain. We measure the raw performance in \emph{millions of cell updates per second} (MCUP/s), so that different problem sizes can be meaningfully plotted side by side.
Since cell updates are an extensive measure, and time-to-solution is not, an additional normalization to a single device size is necessary.  As for the device definition, we typically choose the smallest hardware that can be purchased individually, which may consist of a single or multiple NUMA domains; a notable exception being the PVC hardware (codename from Table \ref{t-dev} for the \emph{Intel Data Center GPU Max 1550}) for which we show results from a single tile, out of its two per card.

In more detail, we perform a weak scaling test until the node is used in its full capacity, then dive by the number of devices per node. Typically, this corresponds to having one MPI task per NUMA domain (or GPU tile). As expected, this configuration is always among those which yield the highest performance; the others are just discarded as poor ways to utilize the node.

\subsection{Energy measurement setup}

We measure our \emph{energy efficiency} in millions of cell updates per Joule of consumed energy (MCUP/J), a metric akin to the FLOPs/W (floating point operations per second per watt of consumed power) used for ranking in the \emph{Green500 list of Supercomputers}.

A difficulty is that energy measuring tools present varied interfaces and usage models that hinder portability and parsing ease of the measure from a user perspective.  Some systems provide global \textit{at-the-plug} measurement, on a \textit{per-batch-job basis}, but these often present user right restrictions (understandably, as energy counters may expose low-level system properties); also they do not necessarily reflect how the actual process is using the hardware, as they depend on all devices connected to the nodes, some of which are site-dependent and may not even be in use (e.g. additional accelerators such as FPGAs, or disks; see \ref{ss-nrg-discusssion}). 

Then, different tools may yield different quantities (energy  consumed in a given time or by a given process rather than average or "instantaneous" power).

Finally, it is hard to eliminate unwanted sections such as initialization and I/O, and results are harder to parse than performance measurement, e.g. compared to built-in performance timers, often making necessary to schedule several jobs to measure statistical uncertainty, wasting resources. 

For this purpose we develop an energy measuring tool, which we distribute alongside DPEcho, focusing on an accessible and portable interface. Given any arbitrary command line tool which prints a low-granularity device power, our shell script (\texttt{deltaEnergy.sh}) parses it to yield the consumed device energy via numerical integration over time bins. Here we use our script from within the internal DPEcho timer, instrumented via an extra thread by the \texttt{Boost} library), taking measures as close to the computing cores as possible. We adopt system management interface (\texttt{smi}) measures for GPUs and (\texttt{perf}, \texttt{likwid} or \texttt{EAR/econtrol} with "cores" or "package" counters for CPUs, never including e.g. RAM or other components) but also standalone usage or custom power meters are possible.  The main ideal requirement is that the provided power meter should have a low latency, considerably smaller to a single simulation wallclock timestep, in order to be statistically accurate.

As \dpecho{} is essentially a device-only USM code, the host energy measures are neglected (but can be quantified too, at will). Users interested in different characterizations of energy efficiency who may want to include broader at-the-plug measurements may do so with trivial modifications. For context, we provide the comparison of the two measures in Section \ref{ss-nrg-discusssion}.

\section{Results}\label{ss-results}

\begin{figure*}[ht]
  \centering
  \includegraphics[width=0.495\linewidth]{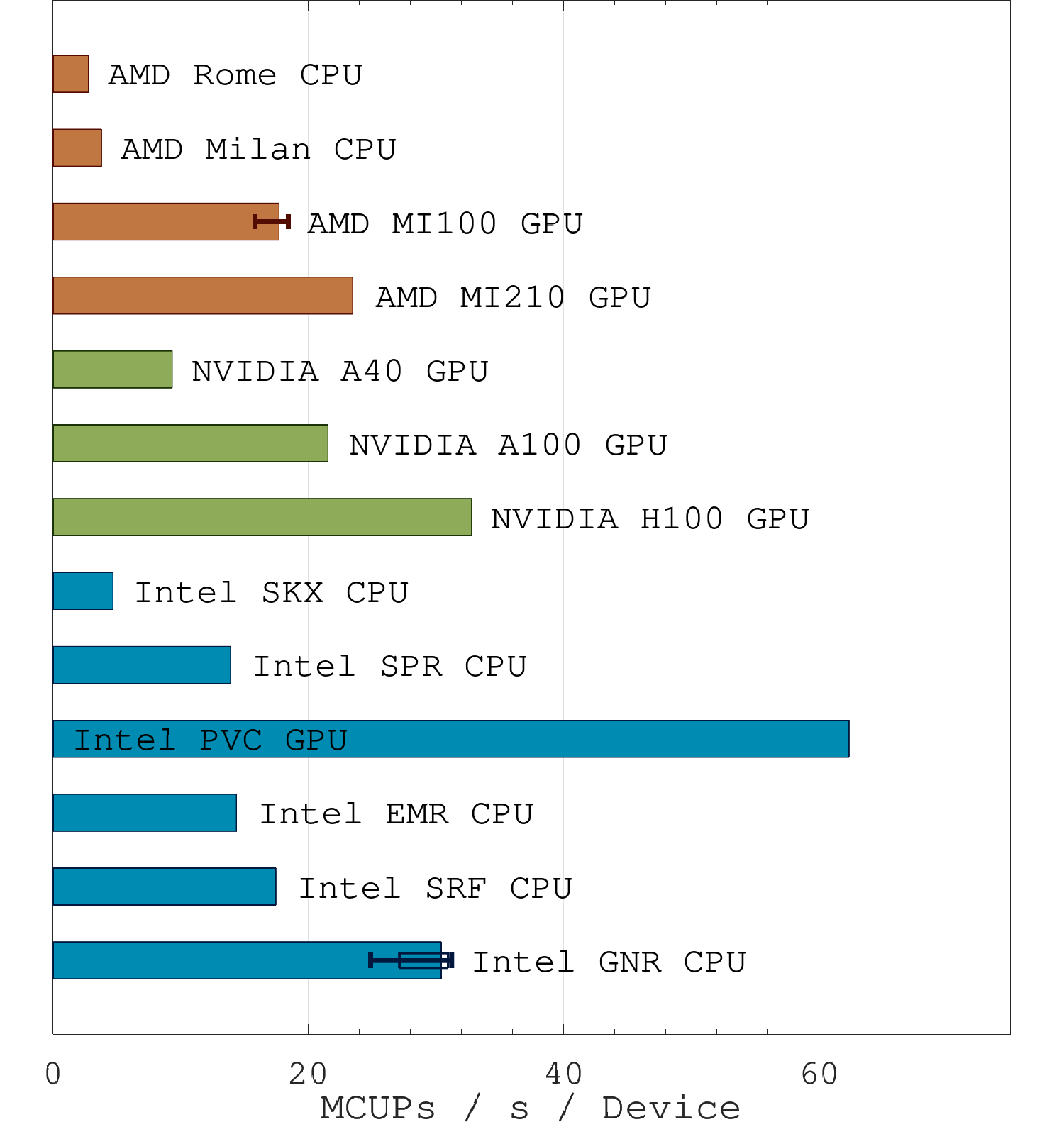}
  \includegraphics[width=0.495\linewidth]{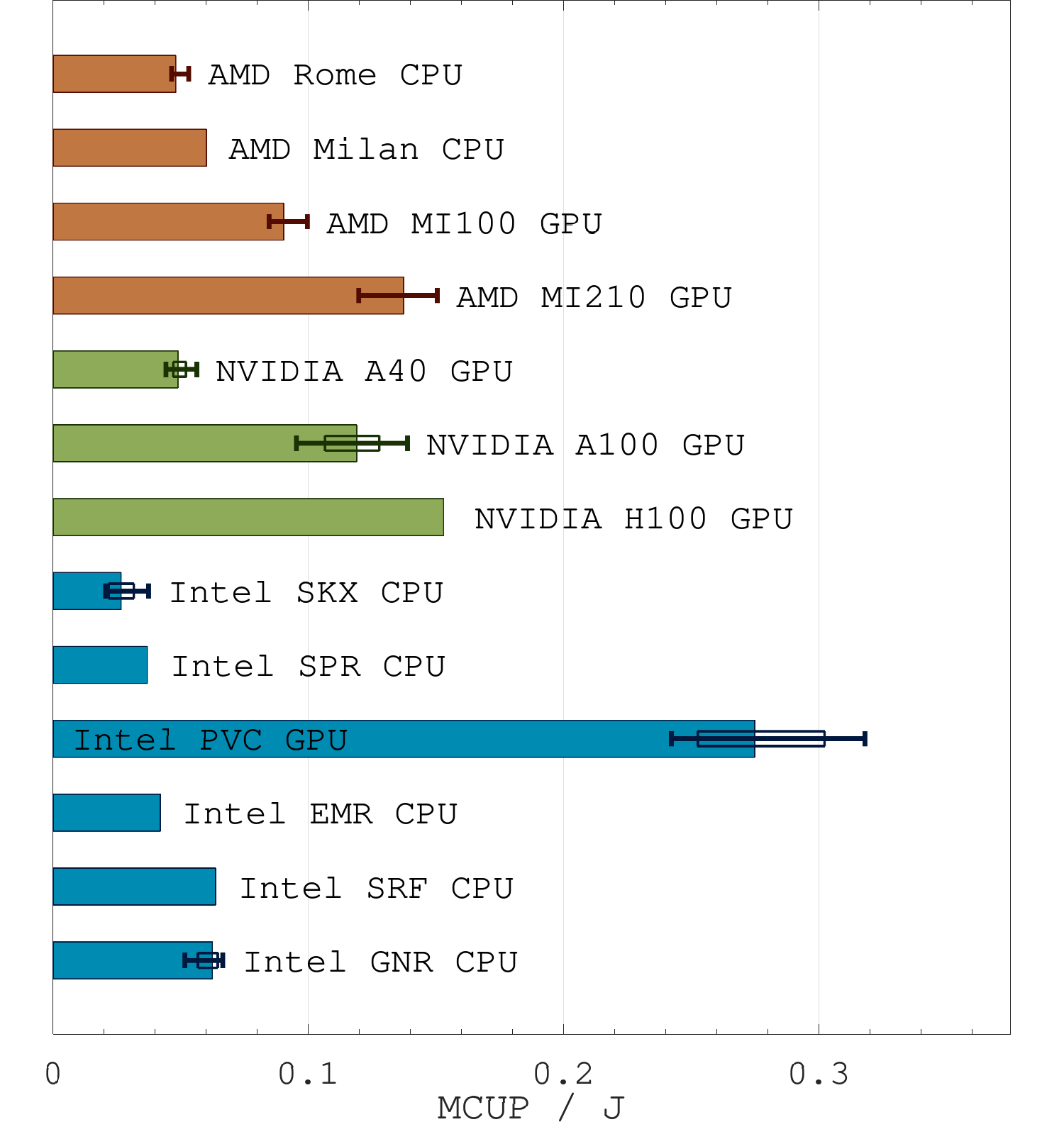}	 
  \caption{\textbf{First panel}: \dpecho{} Performance - MCUPs (millions of cell updates) per second per device, for all hardware listed in Table \ref{t-dev}.
  \textbf{Second panel}: \dpecho{} Energy. - MCUPs per Joule. Note that the normalization over device is not necessary when using this KPI. Data statistics and errors: Median of 21 measurements. Errors represent quartiles (box) and 5th and 95th percentiles (bars). Error bars of either kind smaller than 5\% are not drawn for clarity. Colors highlight different vendors.} 
  \label{f-perf}  
\end{figure*}

Our benchmark results are plotted in Figure \ref{f-perf}, for all hardware presented in Table \ref{t-dev}. For this plot we used the Intel \oneapi{} compilers, alone or with the plugin provided by \emph{Codeplay}\footnote{\url{https://codeplay.com/portal/blogs/2022/12/16/bringing-nvidia-and-amd-support-to-oneapi.html}} for
NVIDIA and AMD GPUs  (which provide a CUDA and HIP backend, respectively). A number of general observations can be made, aside from GPUs largely outperforming CPUs:

\begin{description}
  \item[Performance and energy efficiency differ] - albeit more quantitatively than qualitatively: the relative ranking of the GPU changes little, but their difference with CPUs are exacerbated when looking at energy efficiency.
  \item[GPU performance exhibit large variability] - PVC largely outperforms the other cards but this effect is probably enhanced by hardware-specific optimizations of the Intel compiler (see below; also remember that the PVC performance value refer to a single tile). 
  \item[CPU performance is closely tied to power draw] - as CPU energy efficiencies show much smaller variabilities than their absolute performance.  Intel CPUs show a clear improvement trend with successive generations
  (SRF/GNR $>$ EMR/SPR $>$ SKX), while AMD CPUs show appealing energy efficiencies, in which node power-capping likely plays a role. 
\end{description}
The latter point highlights that looking purely at performance may give an improper view of generational progress, as, in some cases large performance gained are accompanied by greatly increased power draw.
  
The absolute performance per device is systematically lower than in \cite{dpecho2023cielo}, where it peaked around $100\ \mathrm{MCUP / s / Device}$; the reason for this is twofold:
\begin{enumerate}
    \item The previous benchmark referred to a single-device run, having switched off MPI parallelization; this removed MPI overhead, and did not account for the different devices competing for node resources. Being an idealized, non-scalable configuration, it is less representative.
    \item The previous code version presented a simplified flux scheme, correct only with periodic boundary conditions, not allowing for an additional flux grid point in each direction. After the implementation of the correct scheme, extensive tuning of the code was necessary to ensure portability.
\end{enumerate}
The second reason is also tied to why non-intel GPUs have since lost performance relative to PVC, while in literature, for instance, A100 showed par performance with a single PVC tile: the new scheme presents uneven flux grid sizes (e.g. $512\times512\times513$ instead of $512^3$), over which compiler and runtimes may struggle to optimize load sharing within the device. The simplified flux scheme is desirable when running with periodic boundaries, and could be restored provided it does not hinder code maintainability.

\begin{figure*} \centering
  \includegraphics[width=0.495\linewidth]{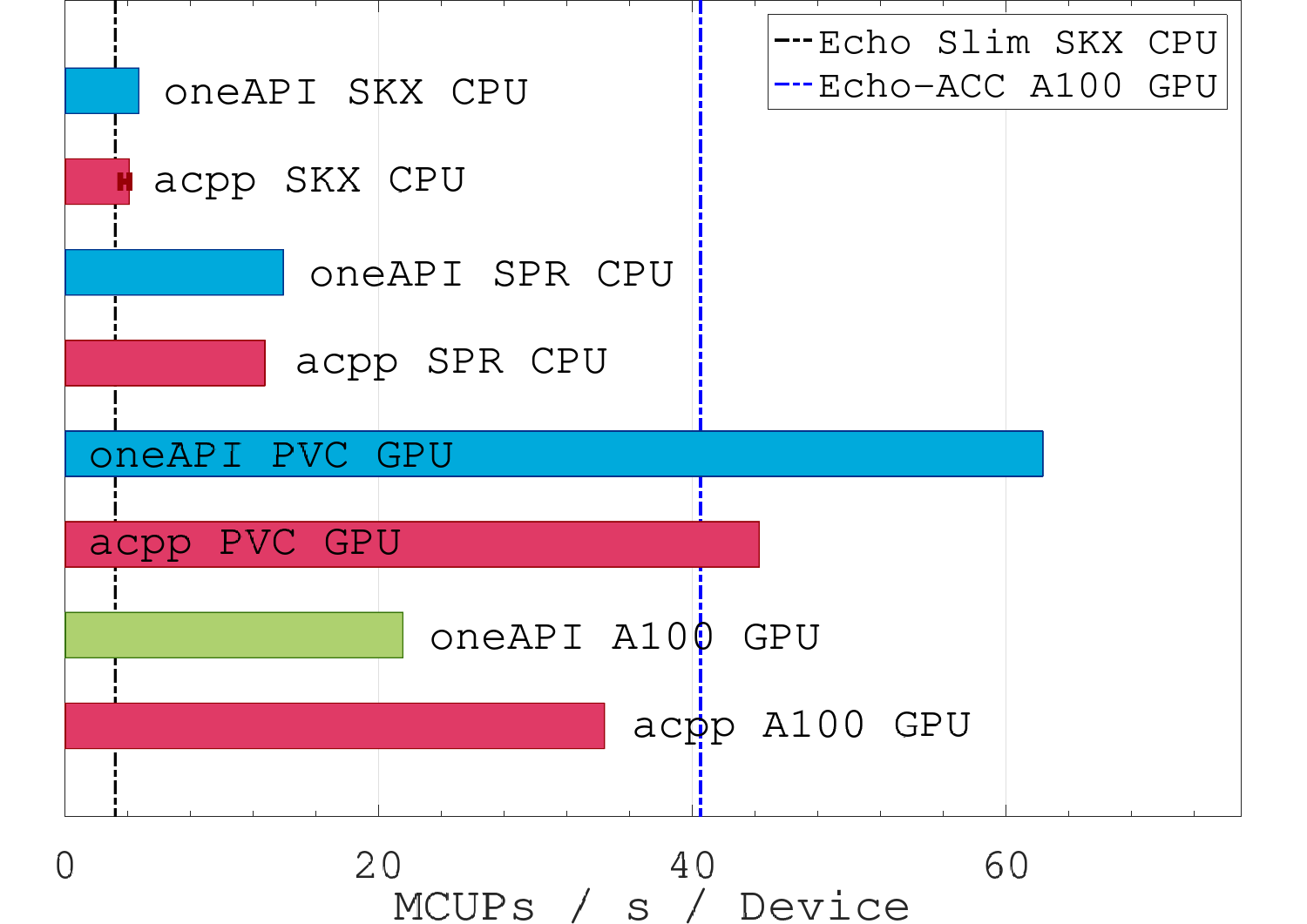} 
  \includegraphics[width=0.495\linewidth]{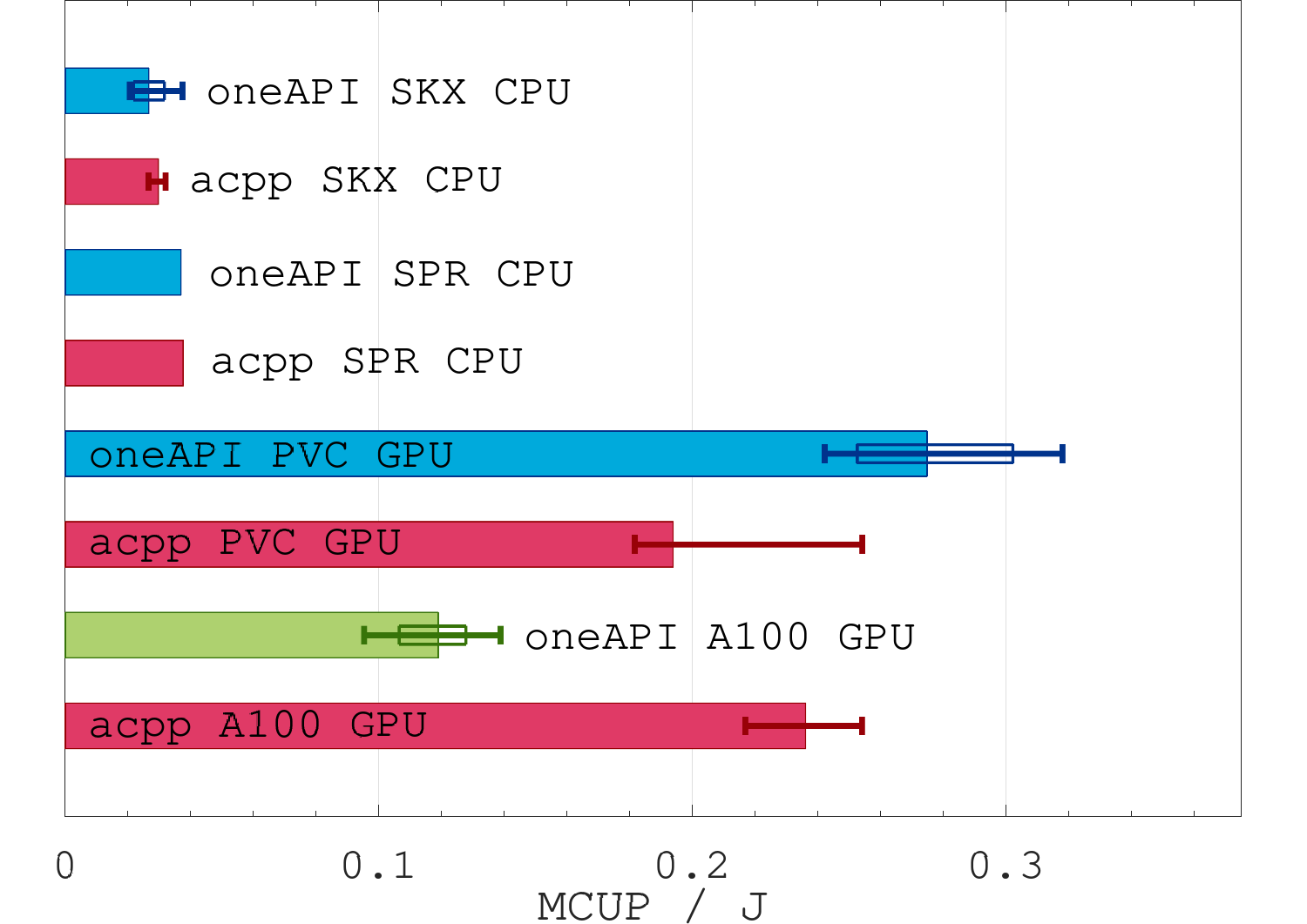}
   \caption{Similar to Figure \ref{f-perf}, but comparing Intel \oneapi{} compiler \texttt{icpx} with AdaptiveCpp's \texttt{acpp} single-pass compiler (for just SKX, SPR, PVC and A100 from Table \ref{t-dev}). For the performance histograms (first panel), we add some indicative lower and upper bounds from public literature (see legend, and text). Intel compilers data are colored according to vendors, \texttt{acpp} data are shown in magenta.
  }
  \label{f-acpp}
\end{figure*}

As a first step in the investigation of these results, we switch to \emph{AdaptiveCpp's} \texttt{acpp} single-pass compiler, in the hope of addressing the above GPU performance issues.
In Figure \ref{f-acpp} we compare the KPIs from \acpp{} with the values from Figure \ref{f-perf}  for the SKX, SPR, PVC and A100 devices.

CPUs show only minimal difference, \acpp{} being marginally lower in performance, yet similar in energy efficiency.  GPUs show a much more dramatic difference; it is clear that using \acpp{} on A100 cards is recommended for \dpecho{}; indeed as recommended in \cite{acpp2023syclcon} it is always good practice to try both compilers, as different choices in the implementation may result in large performance differences. It would be very desirable to extend these benchmarks to other SYCL astrophysical applications as they emerge, and publish the results on a common hub.

\section{Discussion} \label{s-discussion}

\subsection{Comparison with the HPL benchmark} \label{ss-nrg-hpl}

Recent \hpl{} rankings have shown Intel GPUs to be less efficient compared to their competition, in open disagreement with our findings so far.  We thus contrast our energy measures to those of \textit{High-Performance LINPACK} (\hpl{}, \citealp{hpl2003dongarra,hpl2016dongarra}), the most known supercomputer benchmark, used to compile the Top500 and Green500 rankings.

\begin{figure*}
\centering
  \includegraphics[width=0.495\linewidth]{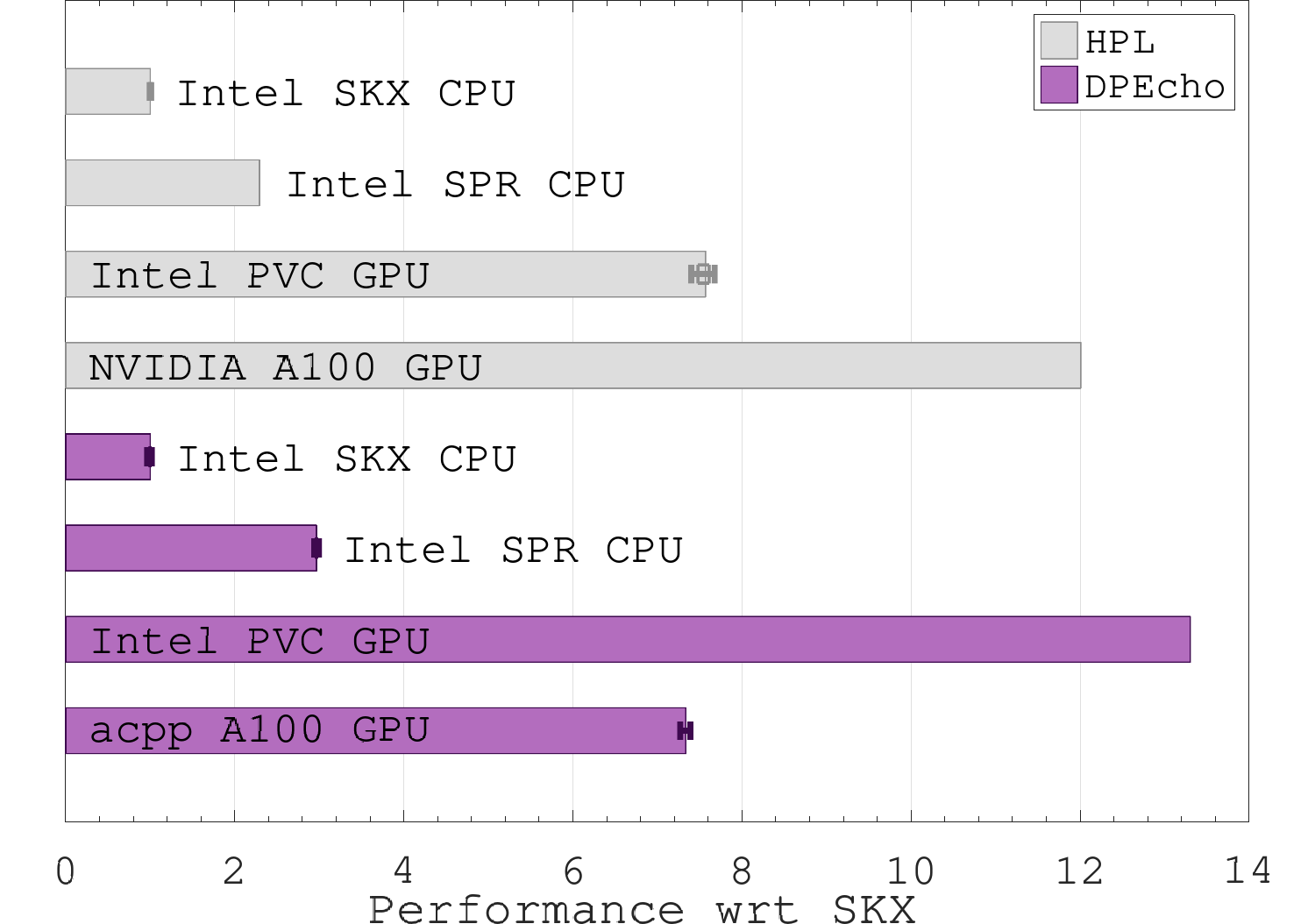}
  \includegraphics[width=0.495\linewidth]{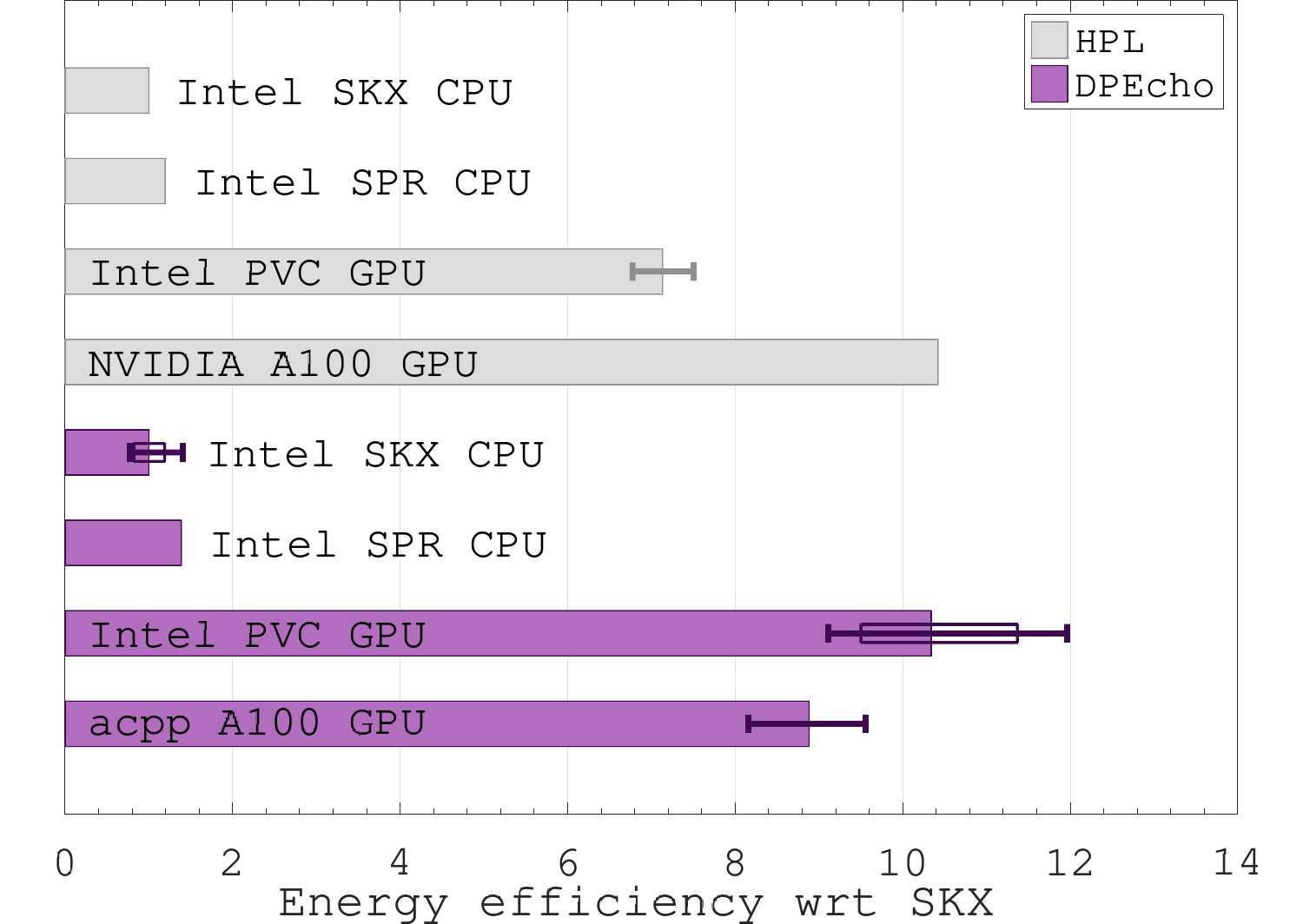}
  \caption{ Comparing performance (\textbf{first panel}) and energy efficiency (\textbf{second panel}) for \hpl{} (in grey) and \dpecho{} (in magenta). }
  \label{f-compared}
\end{figure*}

In Figure \ref{f-compared} we put our results for \hpl{} and \dpecho{} side by side, the underlying HPL metric being GFLOPs/watt; we normalize both  sets by their respective SKX baseline for homogeneity. For \dpecho{} on A100, we used the \acpp{} results.

The comparison clearly highlights the discrepancy: compared to CPUs, PVC grants a better speed-up on \dpecho{}, A100 on \hpl{} (values remarkably close to each other). Thus the two device are comparable as shown in \citealp{dpecho2023cielo} for both metrics (reminding that for PVC we consider a half a card) and once again looking at energy efficiency reduces the difference. 


The \hpl{} authors\cite{hplproblem2016dongarra} warn that \hpl{} is biased towards compute-intensive matrix-multiplication-like (MM) kernels (Type I) over more mixed but potentially more representative applications (Type II), thus recommend supplementing \hpl{} with other metrics. In GPUs the difference could be even larger as, generally speaking, Type I and II kernels target different GPU subsystems (tensor versus vector cores), probing entirely different hardware capabilities. We conclude that \emph{there is no tension between our measurement and the Green500, and reaffirm the superior consistency of energy versus raw performance metric}.

\subsection{Energy of heterogenous hardware} \label{ss-nrg-discusssion}

\begin{SCfigure}
\centering
  \includegraphics[width=0.45\textwidth]{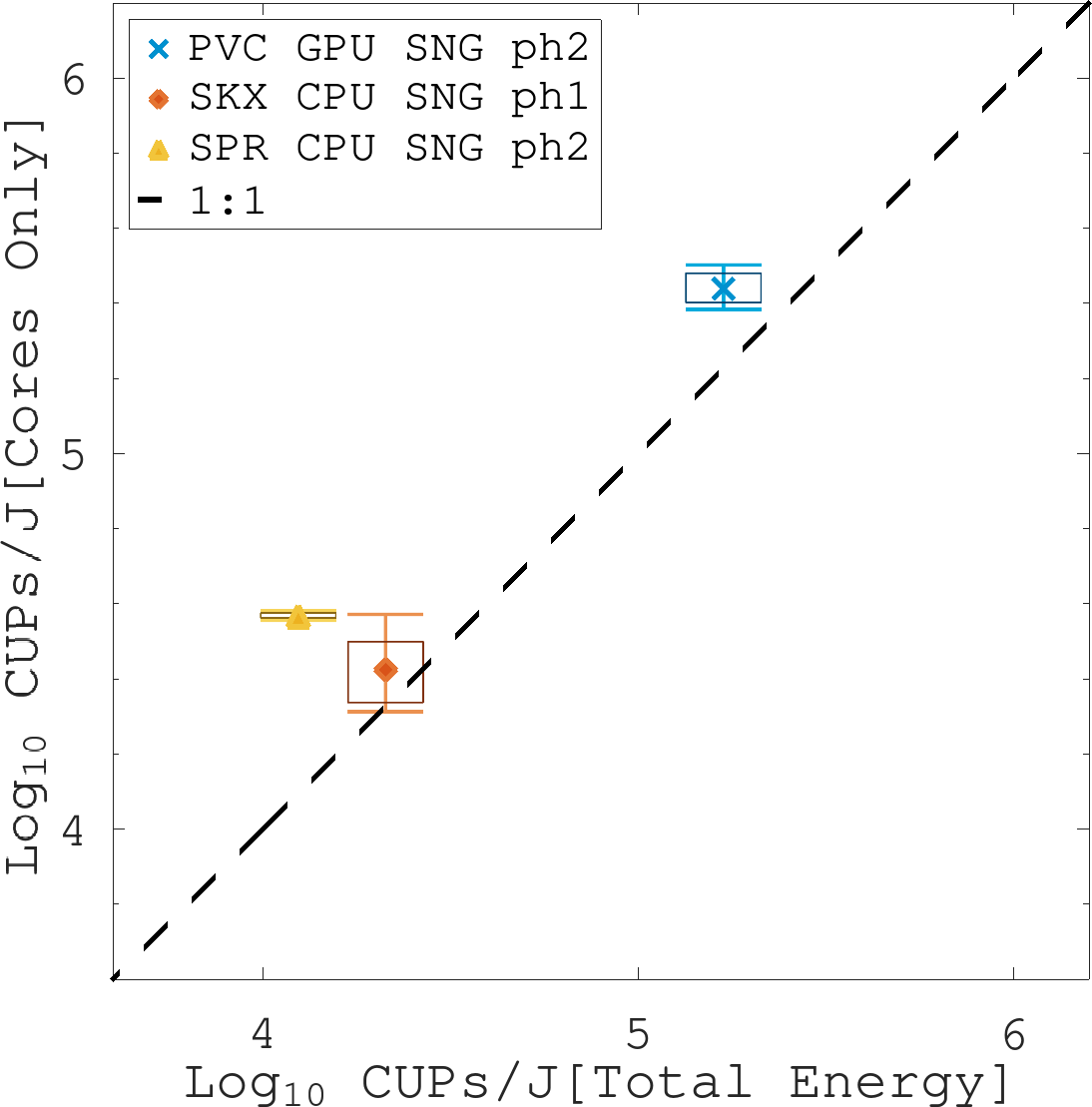}

    \caption{Energy comparison between cores/package and plug node energy. We compare total node DC power consumption from EAR-user-db (x-axis) versus \dpecho{}'s internal energy measure, which targets only GPU/CPU compute cores with hardware-specific methods (y axis, always higher). Besides the x-measure being more impractical, its reproducibility is tied to the exact node composition, and is thus not representative of the compute hardware (see text).
  }
  \label{f-ear}
\end{SCfigure}

\texttt{EAR} (for \emph{Energy Aware Runtime}, \citealp{ear2020corbalan}) is an energy management framework capable of energy accounting, control, and optimization, available for users of SuperMUC-NG  \citep{ear2022lrz} via integration with SLURM. We use \texttt{EAR} to compare quantitatively \dpecho{}'s energy measures, based on cores or packages alone, to a global measure \textit{at-the-plug}. In Figure \ref{f-ear} we plot how the two quantities correlate. The distance of the data points from the given identity diagonal is indicative of the fraction of the energy not spent in compute cores, and depends on the details of the hardware.

As anticipated above, measuring energy on a global level presents a number of issues:
first, the measure is affected by a potential initialization bias, of which we get rid by subtracting the energy from a control run.
Next, we observe large deviations for SPR.
At this time, we only have access to an SPR node that also contains four PVC GPUs, which contribute to the power consumption and do not necessarily stay completely idle during the run, even when just the CPU is targeted. For instance, all devices may be interrogated by the runtime at the moment of device selection.
Both the global approach and our more device-centric approach have merit. 
We chose a device-centric approach for this paper to be able to compare a wide range of compute devices without needing to take the overall node configuration into account.
A global view on energy consumption may be preferable when comparing a small number of specific node configurations, but our focus on core energy allows for more direct device comparison.
Such a discussion on heterogeneous devices triggers a number of observations:

\begin{description}
    \item[Pay attention to the node composition] while benchmarking a particular hardware, especially since users -or even support teams- may have little control on it. Luckily, SYCL offers a comprehensive view of all the installed devices, including CPUs, simply via \texttt{sycl-ls} or \texttt{acpp-info}, albeit the visible devices may depend on backend availability.
    
  \item[Using heterogeneous devices introduces load imbalance] so that it may be advisable to leave lesser devices unused. This can easily achieved with device pinning (automatic like in Intel MPI, or manual 
  such as the round-robin scheme distributed with \dpecho{}), without performance drawbacks. 



\end{description}

\section{Summary and Outlook} \label{s-end}

In this paper, we proposed the use of energy consumption as the primary metric for comparing different devices, due to the intensive quality of the metric, that removes the intrinsic bias of device size.  
DPEcho, an MPI+SYCL proxy application from the astrophysics domain was run on a wide variety of GPU and CPU hardware. The use of SYCL was shown to be both portable and efficient over several devices. Compared to CPUs, the parallel nature of GPU accelerators show a significant advantage in terms of cell updates per energy consumed. Comparison with the HPL benchmark highlighted that energy efficiency still depends on the application being executed. 
These results show encouraging potential for SYCL in computational astrophysics.
 
In future work, we plan to
evaluate the performance of 
\dpecho{} on a variety of hardware platforms, including ARM CPUs, upcoming GPUs, and other emerging commercial hardware. In addition, we encourage the use of DPEcho as a HPC benchmark for energy efficiency,
welcoming readers who wish to run and share their results, which we may include for display on the main \dpecho{} repository with authorship information.
Finally, we plan to instrument additional computational codes with DPEcho’s energy timer to facilitate comprehensive energy consumption analysis.

\ifthenelse{\boolean{doubleblind}}{
\ack
Redacted Redacted Redacted Redacted Redacted Redacted Redacted Redacted Redacted Redacted Redacted Redacted Redacted Redacted Redacted Redacted Redacted Redacted Redacted Redacted Redacted Redacted 
Redacted Redacted Redacted Redacted Redacted Redacted Redacted Redacted Redacted Redacted 

\discl
Redacted Redacted Redacted Redacted Redacted Redacted Redacted Redacted Redacted Redacted Redacted 

Redacted Redacted Redacted Redacted Redacted Redacted Redacted Redacted Redacted Redacted Redacted
Redacted Redacted Redacted Redacted Redacted Redacted Redacted Redacted Redacted Redacted Redacted Redacted Redacted Redacted 
Redacted Redacted Redacted Redacted 
Redacted Redacted Redacted Redacted Redacted Redacted Redacted Redacted Redacted 
Redacted Redacted Redacted Redacted Redacted Redacted Redacted Redacted Redacted 
Redacted Redacted 
Redacted Redacted Redacted Redacted Redacted Redacted Redacted Redacted Redacted Redacted Redacted Redacted 
Redacted Redacted Redacted Redacted Redacted Redacted Redacted 

}{
\ack
SC wishes to thank his colleagues Carla Guillen, Saheed Bolarinwa, Josef Weidendorfer and Amir Raoofy for the support on compilers and energy measurements on SuperMUC-NG phase 1 and 2, and on B.E.A.S.T.
SC also thanks Florian Janetzko and Jolanta Zjupa for making GCS/JSC data available.

\discl
The authors have no competing interests to declare that are relevant to the content of this article. 

Performance varies by use, configuration and other factors. Learn more on the Performance Index site.
Performance results are based on testing as of dates shown in configurations and may not reflect all publicly available updates.
See backup for configuration details.
No product or component can be absolutely secure. Your costs and results may vary.
Intel technologies may require enabled hardware, software or service activation.
© Intel Corporation.
Intel, the Intel logo, and other Intel marks are trademarks of Intel Corporation or its subsidiaries.
Other names and brands may be claimed as the property of others. 
}



\appendix

\section{Programming models for computational astrophysics}\label{a-portings}

Computational astrophysicists are proficient at implementing physical algorithms (e.g. on top of CFD schemes for numerical experiments), which they profitably run at scale (e.g. \citealp{honing2020cielo1}).

For CPU codes, recurrent performance bottlenecks comprise low arithmetic intensity, inefficient memory access patterns, and little use of multithreading (either absent or limited to small kernel scopes).
Scaling problems arise from load imbalance of complex domain  decomposition schemes. 
Well-optimized codes leverage mixed precision operations and efficient CPU vectorization  to produce very large scale, record-breaking simulations (\citealp{modeling2014federrath,beattie2024magnetizedcompressibleturbulencefluctuation,millenniumTNG2022pakmor}, to name a few).
Although they are highly tuned for CPU performance, they end up struggling with GPU compatibility, as new programming paradigms must be introduced.
The refactoring efforts are however complicated by the closed-source nature of several code repositories, which limits community efforts.  
Several online resources exist to get information and code access\footnote{e.g. \url{https://ascl.net/}}, yet in practice, public, active development on state-of-the art codes is limited to the research groups that own the codes.
As for any other code, any useful refactoring requires extensive work and validation; choosing the route of least resistance from the start is paramount.
After having introduced SYCL in the main text, in the following we comment on the advantages and shortcomings of the most common such languages and frameworks for GPU enabling. We provide a visual summary in Table \ref{t-models}, containing useful assessments to plan refactorings, and compatibility for each major vendor (see also the constantly updated table by \citealp{gpucompatibility2023herten}).


\subsubsection*{Fortran/C++}
present inherent parallelism capabilities within the native language most codes are written in (using mainly \texttt{DO CONCURRENT} and \texttt{stdpar} constructs).  Some modern compilers are able to offload regions (e.g. \acpp{}) to accelerators. In practice, even this approach requires extensive code rewriting for algorithm parallelization using the appropriate parallel constructs. This has resulted in successful refactorings (e.g. \citealp{Malenza2024syclperformance, isoCpp2022Lin}); yet a problem with this route is the lack of the concepts of device and shared memory \footnote{\url{https://www.codeproject.com/Articles/5348333/Why-ISO-Cplusplus-Is-Not-Enough-for-Heterogeneous}}, with consequent frequent data non-localities and copies. Nevertheless, offload abstractions without explicit device programming is a very desirable feature.
        
\subsubsection*{Vendor-specific extensions} such  (\texttt{CUDA}, \texttt{OpenACC}, \texttt{HIP}) are being used to effectively fill these shortcomings (e.g. \citealp{delzanna2024echoGPU,gpluto2024order}), and come in both Fortran and C/C++ version. Drawbacks here concern mainly portability outside the vendor hardware; codes from numerical sciences (e.g. \citealp{hemelbGPU2021shealy,hemelb-codes2022hemelb,gromacsArxivAMD2024Alekseenko,gromacsSycl2021Alekseenko}) mend this problem by maintaining multiple code versions, or creating abstract unified interfaces, at the price of continued team efforts at scale. Finally such extensions do not necessarily expose CPUs as devices, a desirable fact in the case of CPUs with high SIMD capabilities, causing further portability problem.
         
\subsubsection*{OpenMP} offload presents arguably the lowest entry barrier of all in terms of code edits, and may target device or shared memory with dedicated allocation constructs (natively or via vendor extensions
\footnote{e.g. \url{https://www.intel.com/content/www/us/en/developer/articles/technical/openmp-features-and-extensions-supported-in-icx.html\#inpage-nav-5}}).
While several astrophysics codes contain already OpenMP threading regions, unfortunately CPU threading and GPU offload require different pragmas; some general parallelism concepts (thread safety, variable visibility in local memory) are valid in both paradigms, but still multiple code versions may be needed, and kernels identified as profitable for threading are not necessarily profitable for offload.
Finally, the offload syntax leaves detailed data mappings to the programmer effort, giving back little control in return; parallel target regions often are bound to follow kernels scopes, again incurring in pricy repeated data transfers.
        
\subsubsection*{Scalable middleware}   such as \emph{kokkos} (\citealp{kokkos2022exa}) and \emph{AMReX} (\citealp{AMReX_JOSS}) introduce common abstract APIs capable on top of lower-level backends. They are indeed among the best solutions, especially as they also synergize with scalable libraries (e.g. AMReX's massively parallel adaptive mesh implementation and \texttt{kokkos-kernels} for linear algebra and graph operations.
They however have a steep learning curve: they do not dispense users from writing kernels with explicit parallelism, may require extensive dedicated support to their API layer, present long developing times, and may not support all backends equally well (e.g. active code such as those presented in \citealp{stone2024athenakperformanceportableversionathena} or \citealp{amrexAstro2010almgren}, still struggle with SYCL backend support at the time of writing). Finally, not all may be viable for Fortran users.

\begin{table*}[]
  \newcommand{\None}{\textcolor{lightgray}{N/A}}
  \newcommand{\Low}{\textcolor{DarkGreen}{Low}} 
  \newcommand{\Shallow}{\textcolor{DarkGreen}{Shallow}}  
  \newcommand{\Medium}{\textcolor{orange}{Medium}}  
  \newcommand{\High}{\textcolor{red}{High}} 
  \newcommand{\Highp}{\textcolor{DarkGreen}{High}}  
  \newcommand{\Lowp}{\textcolor{red}{Low}}  
  \newcommand{\Steep}{\textcolor{red}{Steep}}
  
  \newcommand{\X}{\textcolor{lightgray}{N/A}}
  \newcommand{\Y}{\textcolor{DarkGreen}{\, \ding{51}}}  
  \newcommand{\Maybe}{\textcolor{orange}{(\ding{51})}}  
  \newcommand{\N}{\textcolor{red}{\, \ding{55}}}
  \resizebox{\textwidth}{!}{
\begin{tabular}{m{2.5cm}m{1.2cm}m{1.2cm}m{1.2cm}m{1.2cm}m{1.2cm}m{1.2cm}m{1.2cm}m{1.2cm}m{1.2cm}}
\hline
Paradigm                   & \multicolumn{2}{c}{Language-integrated} & \multicolumn{3}{c}{Language-Extension} & \multicolumn{2}{c}{Directive-based} & \multicolumn{2}{c}{Domain-specific Framework} \\
Language            & C++ std::par   & Fortran do-concurrent  & CUDA  & HIP   & SYCL        & OpenMP    & OpenACC    & Kokkos        & AMREX                \\
\hline
Learning Curve     & \Shallow            & \Shallow       & \Steep & \Steep & \Shallow{} USM  & \Medium   & \Medium    & \Steep         & \Steep                 \\
Refactoring Effort & \Medium         & \Low                   & \High & \High & \Medium     & \Low      & \Low       & \High         & \High                 \\
Control over Data Locality & \None   & \None                  & \Highp & \Highp & \Highp       & \Lowp      & \Lowp       & \Highp         & \Highp                 \\
Compiling Effort           & \Low    & \Low                   & \Low  & \Medium & \High     & \Low      & \Low       & \Medium       & \Medium               \\
Manual Tuning needed       & \None   & \None                  & \Low  & \Low    & \Low{}    & \Medium   & \Medium    & \Low          & \High                 \\
Available for \mbox{Fortran} & \X    & \Y                     & \Y  & \Maybe{} Wrapped & \Maybe{} Wrapped & \Y & \Y & \N        & \Y           \\
Advanced \mbox{libraries}    & \Maybe{} C++26   & \X          & \Y    & \Y     & \Y           & \N       &    \Maybe{}              & \Y                       & \Y         \\           
\hline
Single-source code runs on: \\
\hline
CPUs               & \Y              & \Y                      & \N         & \N            & \Y           & \Maybe                  & \N                 &  \Y                      &  \Y                    \\
NVIDIA GPUs        & \Y              & \Y                      & \Y         & \Y            & \Y           & \Y                 & \Y                  & \Y                       & \Y                     \\
AMD GPUs           & \N              & \N                      & \N         & \Y            & \Y           & \Y                 &  \N                & \Y            & \Y                     \\
Intel GPUs         & \N              & \Y                      & \N         & \Maybe      & \Y           &   \Y               &  \N               &  \Maybe                      & \Maybe                     \\
\hline
\end{tabular}
}\label{t-models}
\end{table*}

\subsection*{SYCL}
        
\textbf{SYCL}\label{ss-sycl} presents a vendor-neutral, multiplatform framework, including also free and open source compilers (\acpp{}, Intel LLVM) which can leverage on different backends in order to expose all the parallelism degrees of most CPUs and GPUs\footnote{Contrary to a common misconception among numerical astrophysicists, SYCL is not specific to Intel hardware}. As an extension of C++, SYCL is built on specific standards, which it complements with constructs for buffered or unified memory between host and device.  Also noteworthy are the publicly available CUDA-to-SYCL conversion tools: \emph{DPCT}\footnote{\url{https://www.intel.com/content/www/us/en/developer/tools/oneapi/dpc-compatibility-tool.html}}, and its open source counterpart \emph{SYCLomatic}\footnote{\url{https://github.com/oneapi-src/SYCLomatic}} (\citealp{syclomatic2022moffat}). These automatically produce executable SYCL code that can be refined into more idiomatic SYCL and tuned for performance, or ask for programmer input if no direct CUDA-SYCL match is found. This is typically sufficient as an initial baseline, and to get a working port quickly\footnote{Related: the \emph{Intel Application Migration Tool for OpenACC to OpenMP API}\citep{servat2024potconversion}, used, e.g., to port POT3D to OpenMP: \url{https://github.com/UCBoulder/GEM_Intel}}.

An important drawback of SYCL is its unavailability for Fortran; thus Fortran users may want to look at other offload strategies. SYCL also lacks extensive algebra and mathematical libraries, but the gap can be filled with third-party libraries such as Intel oneMKL. Aside from this cases, Astrophysicists who choose SYCL for porting largely benefit from the portability, being able to apply for compute time on virtually every system. SYCL is already behind several such refactorings, in both astrophysics\cite{idefix2023lesur,kashino2022multi,rangel2023performance,malenza2024performance} and other numerical sciences\cite{gromacsSycl2021Alekseenko,hemelbGPU2021shealy}.

\section{Drawbacks met while developing \dpecho{}}
We would now like to share the insight we matured by working on the SYCL benchmark \dpecho{}, in the hope of promoting best practices for SYCL programming, to help avoiding pitfalls to readers interested in SYCL refactorings, and presenting the advantages of SYCL for projects that stagnate from lack of opportunity, convincing alternatives, and human/time resources. 

With the public code \dpecho{}, available for interested astrophysicists to test and develop, we are capable of benchmarking most HPC-grade hardware with a single-source realistic workload. We remind that \dpecho{} is a SYCL rewrite of Fortran90-based GR-MHD, fixed-grid code ECHO \citep{echo2007delzanna} comprising a limited selection of its physics. Thanks to SYCL Unified Shared Memory (USM), data resides in GPU memory for the entire duration of the computation; I/O operations are dealt with one-way \texttt{memcpy} operations.

\subsubsection*{Multidimensional arrays} 

Grid calculations in the Fortran version simply utilize the language support for multi-dimensional array indices, as well as array slices. Unfortunately the USM paradigm only allows for one-dimensional indexing. Some libraries, such as Eigen, or, SYCL's own buffered memory, introduce their own indexing types; but a standardized vocabulary type has only been added with \emph{std::mdspan} in C++-23.  However, at the time of writing, mdspan is not yet supported by all major compiler vendors' standard libraries (specifically, the GNU C++ standard library, which is used by the Intel Compiler), and we were thus not able to use it.
The lack of a unified abstraction gave rise to a multitude of different implementations within applications, such as the \emph{Grid} class within \dpecho{}. 
While these solutions usually fit their respective applications well, they typically necessitate wrappers and adaptors when the application needs to interface with externally developed components. Even when such abstractions are programmed on top of the raw pointers,
in a 3D code in which kernel and indexing are to be written explicitly, this is a tremendous development bottleneck.

\subsubsection*{Compiling} 

When using \dpecho{} for benchmarking the hardware of Table \ref{t-dev}, we were often in the condition of having to install a suitable SYCL compiler by ourselves.

Users of systems mounting Intel CPUs, virtually always finds the \texttt{icpx} compiler and Intel MPI as part of the \oneapi{} toolkits. These are enough to compile SYCL code and run at scale, with even facilitated GPU task pinning \footnote{\url{https://www.intel.com/content/www/us/en/docs/mpi-library/developer-reference-linux/2021-8/gpu-pinning.html}}.

On other systems, generally vendors show no interest in providing or supporting any SYCL compiler, thus users and support teams have to rely on third parties and install compilers themselves; a clearly undesirable situation.

For x86 CPUs, or when the latest compiler version is needed, users can install binaries directly from Intel \oneapi{} toolkits autonomously. Support for NVIDIA and AMD GPUs can be added via plugins by Codeplay\footnote{\url{https://codeplay.com/solutions/oneapi/plugins/}}; the procedure is quick and straightforward, but those require a specific CUDA/ROCM versions. 

In such cases, installation from source is needed. This route is in practice and comprehensibly met with resistance by programmers and data centre support teams, due to the effort and time investment from their part; also for multi-GPU codes, explicit  task-pinning strategies may have to be implemented, implying potential code changes.

Luckily there are several options:
\begin{description}
    \item[the Intel LLVM  compiler]  \footnote{\url{https://github.com/intel/llvm}} provides an open-source and often updated alternative which include AMD and NVIDIA GPU support 
    \item[AdaptiveCpp's \acpp{} compiler] \footnote{\url{https://github.com/AdaptiveCpp/AdaptiveCpp}}is not only a cross-platform single-pass SYCL compiler, but also a general, offload capable, C++ compiler. Its most recent releases feature a virtually complete support for all SYCL2020 specification. Its dependency include an LLVM install; by experience heavy on entry-level user. 
    \item[SPACK] package manager contains recipes for both, largely automating an otherwise complicated process \citep{gamblin2015spack}.
\end{description}

On their side, data center application support teams should -and are invited to- provide or share such install throughout their systems to minimize user effort.

\section{DPEcho as MPI Benchmark}

We finally show the benchmark results at scale. A word of warning: while the full GR-MHD \dpecho{} provides a realistic workload at node (or task) level, the MPI scaling of the given test problem (uniform Alfv\`en waves) is rather idealized, as all ranks work on essentially identical boxes (save for the phase of the wave and the topology shared boundary, which may vary). On the positive side, this allows to benchmark the scaling performance of the hardware with a well-balanced load.

\begin{SCfigure}[h]
  \centering
  \includegraphics[width=0.45\textwidth]{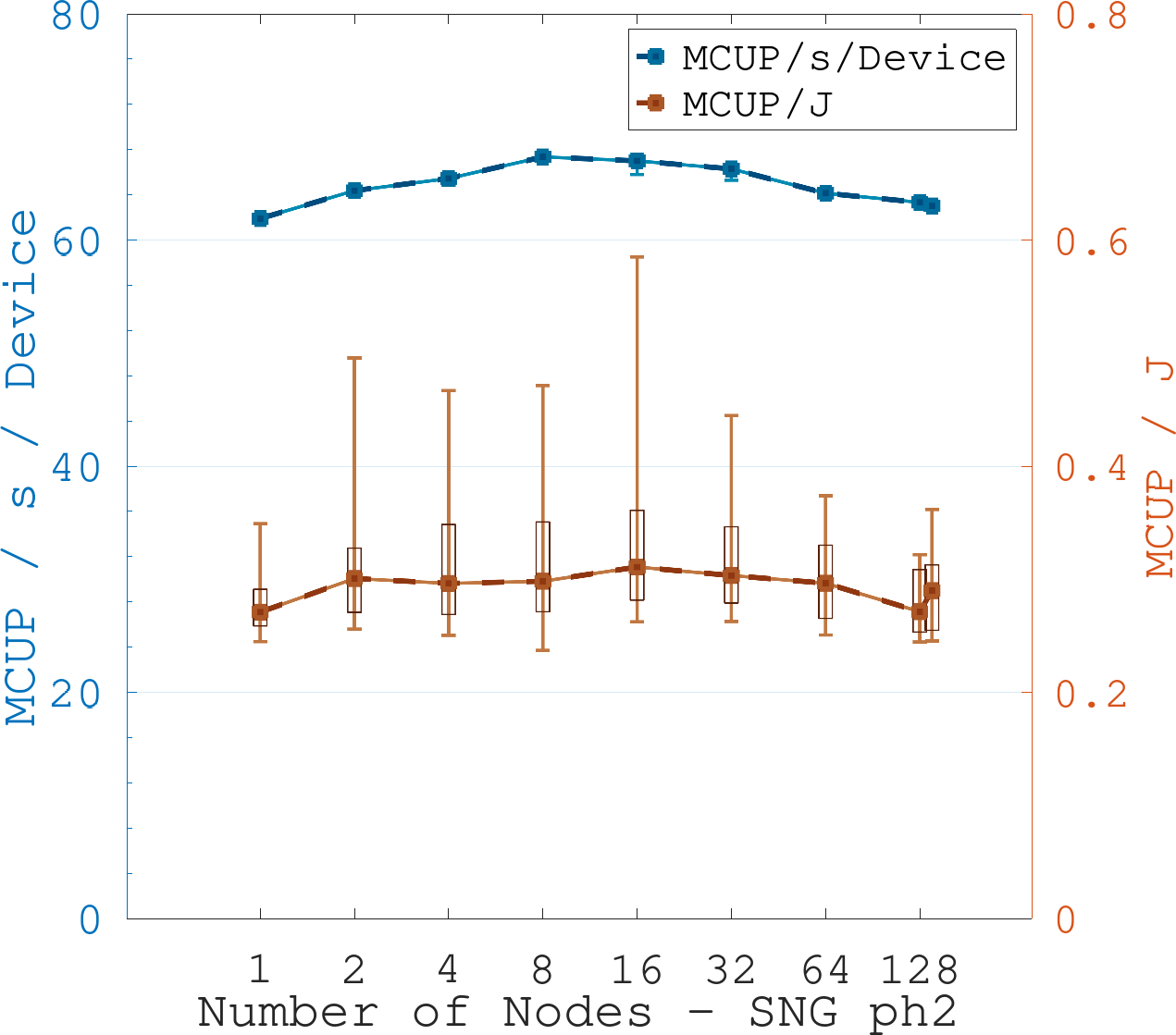}
  \caption{MPI weak scaling of \dpecho{} on \emph{SuperMUC-NG phase2} up to 140 nodes (with 8 PVC tiles each) for both KPIs: performance (in lightblue) and energy efficiency (in orange).}
  \label{f-mpi}
\end{SCfigure}
In Figure \ref{f-mpi} we plot weak scaling results up to 140 nodes of SuperMUC-NG (PVC), in which the number of cells per node is kept constant, and the domain is proportionally increased in resolution and split by enlarging the same MPI Cartesian grid used for the single node test.  

Both the performance and energy efficiency KPIs, aside from the different normalization show the same behaviour; the energy efficiency (in orange) presents artificially high errorbars, due to the high intrinsic latency (1 second being the minimal allowed time granularity) of the underlying power measurement application, \texttt{xpu-smi}. This is a valuable warning in itself, and eventual improvement in this directions are being presented to system administrators and developers of the application.

At large scale, we observe no appreciable decline in performance until 32 or 64 nodes (128 or 256 PVC tiles), and overall scaling efficiency above 90\%. At small scales the performance counter-intuitively increases from 1 to 8 or 16 nodes. This is likely due to specific machine settings, where options that should optimize single-node loads in effect reduce performance (e.g. direct GPU-GPU MPI calls only available intranode, that may not be profitable in all situations); the geometry of the threedimensional binary cartesian decomposition, optimal when the ranks are a multiple of 8, also plays a role in the variability.


\bibliographystyle{elsarticle-harv} 
\bibliography{journals,all}

\end{document}